\documentclass[useAMS,usenatbib]{mn2e}
\usepackage{times}
\usepackage[fleqn]{amsmath}
\usepackage{amssymb}
\usepackage{graphicx}
\usepackage{subfigure}
\usepackage{float}
\usepackage{relsize}
\usepackage{color}
\usepackage{gensymb}
\usepackage[mathscr]{euscript}

\title[Toroidal vortices in protoplanetary discs]{Toroidal vortices and the conglomeration of dust into rings in protoplanetary discs}\author[Pablo Lor\'en-Aguilar and Matthew R. Bate]{Pablo Lor\'en-Aguilar$^{1}$\thanks{E-mail:
pablo@astro.ex.ac.uk} and Matthew R. Bate$^{1}$ \\ $^{1}$ School of Physics and Astronomy, University of Exeter, Stocker Road, Exeter EX4 4QL, United Kingdom}

\date{Submitted for publication to MNRAS}

\begin{document}

\pagerange{\pageref{firstpage}--\pageref{lastpage}} \pubyear{???} \maketitle
\label{firstpage}

\begin{abstract} 
We identify a new hydrodynamical instability in protoplanetary discs that may arise due to variations in the dust-to-gas ratio and may lead to concentration of dust grains within a disc.  The instability can arise due to dust settling, which produces a vertical compositional entropy gradient.  The entropy gradient drives a baroclinic instability that is capable of creating toroidal gas vortices that gather dust into rings.  Such dust rings are potentially observable via continuum emission of the dust or scattered light.  Indeed, this instability may offer an explanation for the rings recently observed in the discs around the young stars HL Tau and TW Hya that does not rely on clearing by protoplanets.   The instability may also have wider ramifications, potentially aiding dust agglomeration, altering the radial migration of larger planetesimals, and modifying angular momentum transport within a disc.
\end{abstract}


\section{Introduction}

During the early phases of planet formation, dust grains are hypothesised to agglomerate into planetesimals.  However, collisions between planetesimals larger than a few centimetres in size are generally thought to lead instead to bouncing or fragmentation \citep{Wei77,WeCu93, BlumWurm2000}. In addition, in laminar gas discs, metre-sized planetesimals are predicted to suffer rapid orbital decay due to gas drag \citep{Wei77}. Various mechanisms have been proposed to concentrate dust and planetesimals at particular locations in discs, either to avoid radial migration or to allow them to grow \citep*[e.g.][]{Saf69,GW73,NSH86,Wei80,BargeSommeria1995, TBD96, Rea04, YG05, JK05}.  In particular, baroclinc instabilities in protoplanetary discs may form large-scale vortices \citep{KB03, Klahr04, PJS07}, and these have been suggested as a possible mechanism to trap dust particles \citep{BargeSommeria1995, TBD96, GodonLivio1999, JAB04, KB06}. In these past studies, the baroclinic instabilities have been proposed to arise due to radial thermal entropy gradients in the disc and produce vortices coplanar with the disc.

A dust particle moving through gas experiences drag. The form of the drag force may vary considerably as a function of the grain and gas properties \citep{Wei77}, but if the mean free path of the gas molecules is larger than the dust particle's radius, the characteristic timescale for the dust particle's speed relative to the gas to decay (the stopping time) depends linearly on its density and size, and inversely on the density of the gas.  Thus, the behaviour of dust particles in a gaseous disc around a young star depends on their size.  Small particles ($\ll 1$~mm for typical protoplanetary disc densities) have short stopping times and are well-coupled to the gas. The dust-gas mixture behaves almost as a single fluid, with a lower pressure for a given density than for pure gas (since the dust does not contribute to the pressure).  Large particles (typically $\gg 10$~m) are poorly coupled and move on almost ballistic trajectories, barely affecting the gas.  Particles with intermediate sizes affect the inertia of the dust-gas fluid, but also migrate relative to the gas.  In particular, if such particles are initially well-mixed in the gaseous disc surrounding a young star, they will tend to slowly settle \citep{GW73} towards the mid-plane of the disc under the action of the vertical component of the gravitational field exerted by the star.\\

Dust settling is a phase separation process. One therefore expects to have a strong entropy gradient at the interface between the settled, gas/dust mixture, and the purer gas phase above.  In a gaseous protoplanetary disc surrounding a young star, settling of dust towards the mid-plane of the disc will produce a vertical gradient of the dust-to-gas ratio above the disc's mid-plane.  In this letter, we show that this vertical compositional entropy gradient may produce an azimuthal baroclinic instability that drives toroidal vortices within the disc.  In Section \ref{instability}, we investigate analytically the criteria for instability.  In Section \ref{numerical}, we describe the numerical method and initial conditions that we use for hydrodynamical simulations of dusty protoplanetary discs.  Results from our calculations are presented in Section \ref{results}, and in Section \ref{conclusions} we give our conclusions. 

\begin{figure} \centering
\includegraphics[width=85mm]{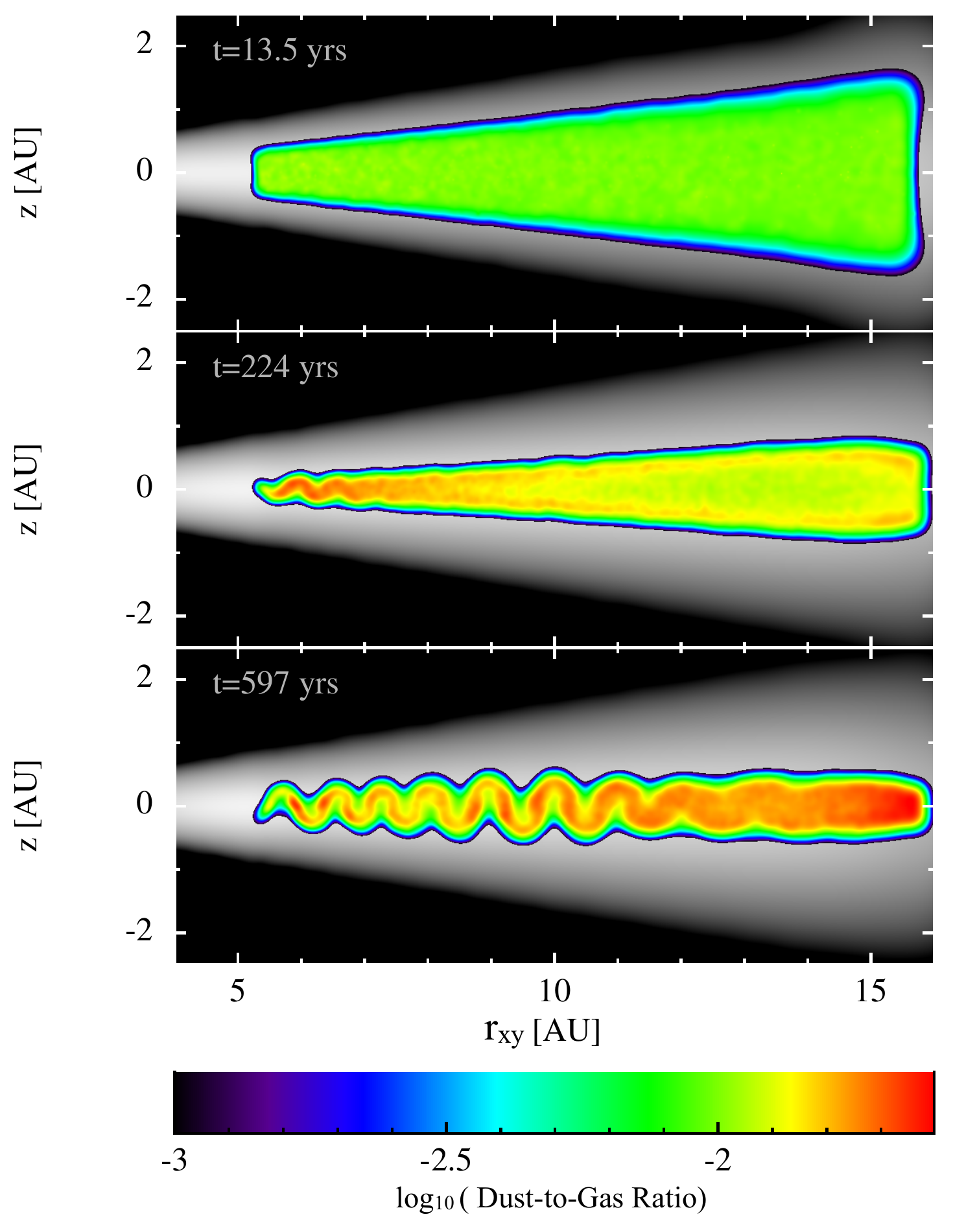}
\caption{The time evolution of the azimuthally-averaged dust-to-gas ratio (colour rendering, maximum $\epsilon\approx 0.025$) and gas density (greyscale, ranging from $\log_{10}(\rho_{\rm gas})=[-16,-11]$) for a locally-isothermal gaseous circumstellar disc with $H/r=0.05$ and an initial dust-to-gas ratio, $\epsilon=0.01$ comprised of 1 mm dust grains. The dust initially settles toward the mid-plane (top panel). After the dust reaches its quasi-stationary state, the compositional baroclinic instability rapidly develops, first at small radii (middle panel), and then throughout the disc (lower panel). The radial wavelength of the instability is approximately equal to the thickness of the quasi-stationary dust layer.  The calculation was performed with $N_{\rm p} = 3\times 10^6$ particles.}
\label{fig:evolution}
\end{figure}
\section{Hydrodynamical instability}
\label{instability}
Baroclinity is described mathematically by the curl of the pressure gradient term in the fluid momentum equation
\begin{equation}
\nabla \times \left({-\frac{1}{\rho}\nabla{P}}\right)= \frac{1}{\rho^2}\nabla{\rho}\times{\nabla P},
\label{barotropic}
\end{equation}
where $P$ and $\rho$ are the fluid pressure and density, respectively. In an axisymmetric dusty gas in which the dust is well coupled to the gas and the mixture behaves as a single fluid, the only non-zero component of equation \ref{barotropic} in cylindrical coordinates is given by
\begin{equation} 
\frac{1}{\rho^2}\nabla{\rho}\times\nabla P 
 = \frac{P}{K\rho^2}\left[\frac{\partial\rho}{\partial z}\frac{\partial K}{\partial r} +\gamma\frac{K}{1+\epsilon}\left(\frac{\partial\rho}{\partial r}\frac{\partial\epsilon}{\partial z} - \frac{\partial \rho}{\partial z}\frac{\partial\epsilon}{\partial r}\right)\right]\hat{e}_{\rm \theta},
\label{axisymmetry}
\end{equation}
\noindent where $\hat{e}_{\rm \theta}$ is the azimuthal unit vector and, for generality, we have assumed a barotropic equation of state given by $P = K\rho_{\rm gas}^\gamma = K\left(\frac{\rho}{1+\epsilon}\right)^\gamma$, 
where $K(r)$, the total density of the fluid $\rho = \rho_{\rm gas} + \rho_{\rm dust}$ is the sum of the densities of the gas and dust, and $\epsilon \equiv \rho_{\rm dust}/\rho_{\rm gas}$ is the dust-to-gas ratio.
\begin{figure*}\centering \vspace{-5pt}
\includegraphics[width=145mm]{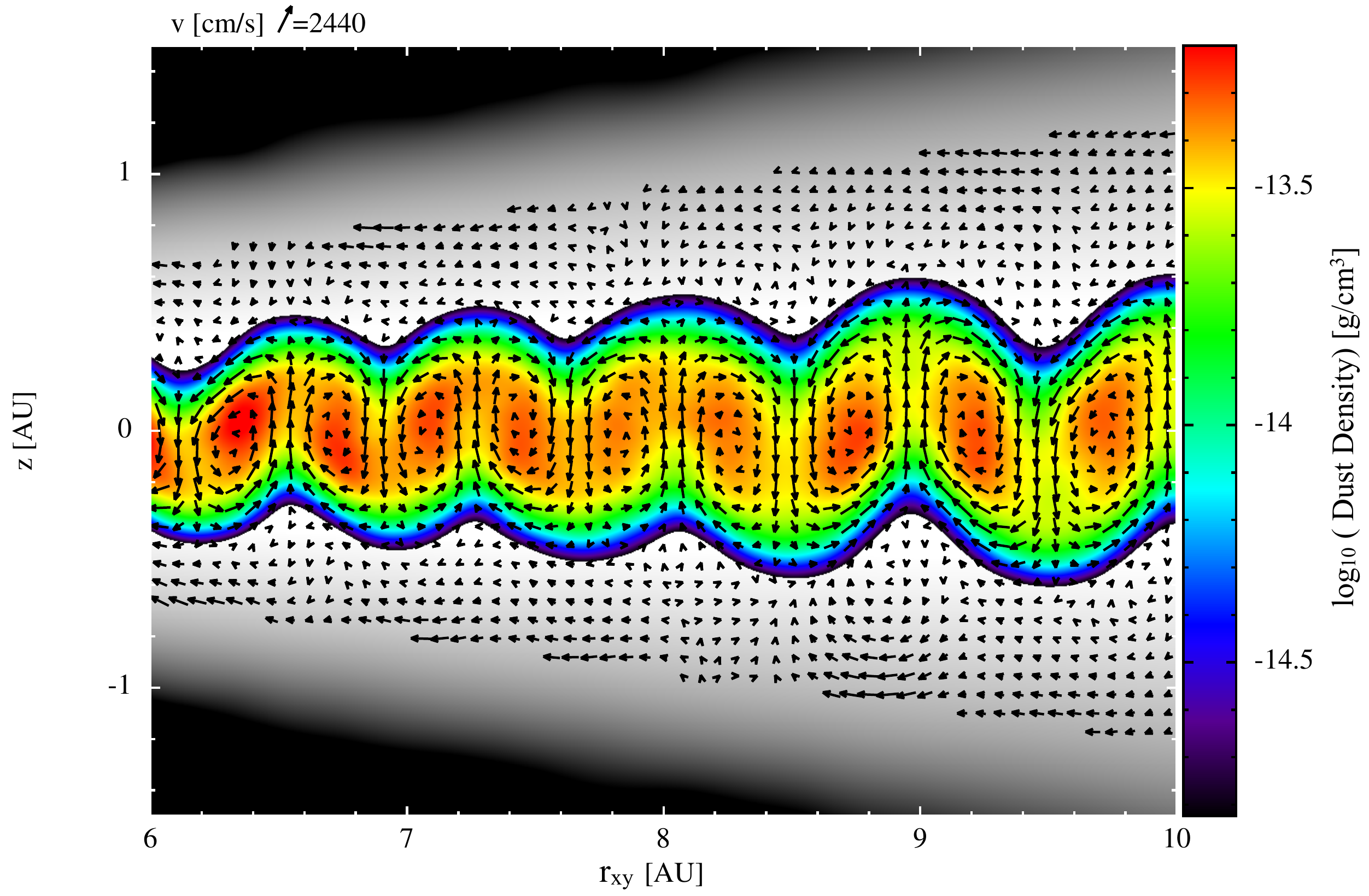} \vspace{-6pt}
\caption{For the same case as shown in the lower panel of Fig. 1, we plot the velocity field (vectors) and the azimuthally-averaged dust density (colour rendering) and gas density (greyscale, ranging from $\log_{10}(\rho_{\rm gas})=[-17,-12]$ in g~cm$^{-3}$). The toroidal vortices within the dust layer are clearly evident, and the dust density is highest at the centres of the vortices (i.e. producing concentric rings).}
\label{fig:vortices}
\end{figure*}
A non-zero baroclinity will change the vorticity, potentially producing vortices. However, the presence of baroclinity by itself does not necessarily lead to instability due to the stabilising effect of rotation. In a typical gaseous accretion disc, the vertical density stratification (first term on the right hand side of equation \ref{axisymmetry}) is not strong enough to make the disc unstable. However, as can be seen in the last term on the right hand side of equation \ref{axisymmetry}, the presence of a compositional gradient may significantly increase baroclinity, potentially making a disc unstable.  As has been beautifully explained in the context of thermally driven baroclinic instabilities  \citep*[e.g.][]{LK11}, by analogy, a compositional instability may be initiated due to the presence of strong gradients in the dust-to-gas ratio, and the resulting baroclinity will generate vorticity. The emerging vortices mix the fluid, moving high entropy well-mixed fluid from the dusty settled layers into the lower entropy purer gas, thus reinforcing the instability. 

The two Solberg-H{\o}iland stability conditions for a gaseous disc  \citep{Tas00} are
\begin{equation} \begin{aligned}
&\frac{1}{r^3}\frac{\partial j^2}{\partial r} - \frac{1}{c_{\rm p}\rho} \nabla P\cdot \nabla S > 0, 
\\
&\partial S/\partial z > 0,
\label{SH}
\end{aligned} \end{equation}
\noindent where $j = r^2\Omega$ is the Keplerian specific angular momentum and $\Omega$ is the Keplerian angular frequency, $c_{\rm p}$ is the specific heat capacity of the gas at constant pressure, and $S$ is the entropy of the fluid. These equations will still be valid in a dusty gas as long as the mixture behaves as a single fluid (i.e. the dust stopping time is short) and the specific heat capacity of the mixture is dominated by that of the gas (i.e. the dust-to-gas ratio is small).

The first of the Solberg-H{\o}iland conditions will be violated if the Brunt-V{\"a}is{\"a}l{\"a} frequency squared (the second term) dominates over the epicyclic frequency squared (the first term).  Due to dust settling, the vertical gradient in the dust-to-gas ratio will initially dominate the radial gradient, so $|\nabla S| \approx |\partial S / \partial z|$ and 
\begin{equation} \begin{aligned}
\label{grad_S}
\frac{\partial S}{\partial z} &= c_{\rm v}\frac{\partial }{\partial z}\log\left(P/\rho_{\rm gas}^{\gamma}\right) = c_{\rm v}\frac{\partial }{\partial z}\log\left(\frac{P}{\rho^{\gamma}}(1+\epsilon)^{\gamma}\right)\\
&=-\gamma c_{\rm v}\left(\frac{\partial\log \rho}{\partial\log P}-\frac{1}{\gamma}\right)\left(\frac{1}{\rho}\frac{\partial P}{\partial z}\right)\frac{\rho}{P} +\gamma c_{\rm v}\frac{\partial}{\partial z} \log(1+\epsilon)\\
&= -\gamma c_{\rm v}\left[\gamma \Delta\nabla g_{\rm z}/c^2_{\rm s} -\frac{\partial}{\partial z} \log(1+\epsilon)\right], \end{aligned} 
 \end{equation}
where \citep{Cab84},
\begin{equation}
\Delta\nabla \equiv \frac{\partial\log \rho}{\partial\log P}-\frac{1}{\gamma}, 
\label{Cabbot}
\end{equation}
$c_{\rm v}$ is the specific heat capacity of the gas at constant volume, the sound speed is $c_{\rm s}=\sqrt{\gamma P/\rho}$,  and $g_{\rm z} = -\Omega^2 z$ is the vertical component of the acceleration due to the star's gravity. The quantity in equation \ref{Cabbot} is usually zero in an equilibrium situation, but in a convectively stable region typically adopts values $\Delta\nabla \approx 0.1$ \citep{Cab84}. In our locally-isothermal disc models discussed below, $\Delta\nabla = 0$, so the vertical entropy gradient arises purely due to the dust-to-gas ratio vertical gradient.  We can then write
\begin{equation} \begin{aligned}
&-\frac{1}{c_{\rm p}\rho}{{\nabla}} P\cdot {\nabla} S \approx -g_{\rm z}\left[-\gamma\Delta\nabla g_{\rm z}/c^2_{\rm s} +\frac{\partial}{\partial z} \log(1+\epsilon)\right]\\
&=\gamma\Delta\nabla\Omega^4z^2/c^2_{\rm s} + \Omega^2z\frac{\partial}{\partial z} \log(1+\epsilon),
\end{aligned} 
\label{eq:BV}
\end{equation}
Therefore, the system is potentially unstable at $z>0$ if 
\begin{equation} 
\begin{aligned}
&\left|\frac{\partial \epsilon}{\partial z}\right|  \gtrsim   \frac{z}{H^2}\left(1+\epsilon\right)\left[\gamma \Delta\nabla+\left(\frac{H}{z}\right)^2\right], \\
&\frac{\partial \epsilon}{\partial z} < 0,
\label{eq:instability} \end{aligned} 
\end{equation}
\noindent where we have defined the vertical scale height of the disc $H = c_{\rm s}/\Omega$. The former equation implies that it becomes harder for the instability criterion to be satisfied for small $z$, and the most favourable location for the development of the instability  is
\begin{equation}
\frac{z}{H} = \frac{1}{\sqrt{\gamma\Delta\nabla}}  \implies \left|\frac{\partial \epsilon}{\partial z}\right| > \frac{2\left(1+\epsilon\right)\sqrt{\gamma \Delta\nabla}}{H}.
\end{equation}
If $\Delta\nabla = 0$, the criterion for instability is more easily satisfied for increasing $z$.

\begin{figure}\centering
\includegraphics[width=80mm]{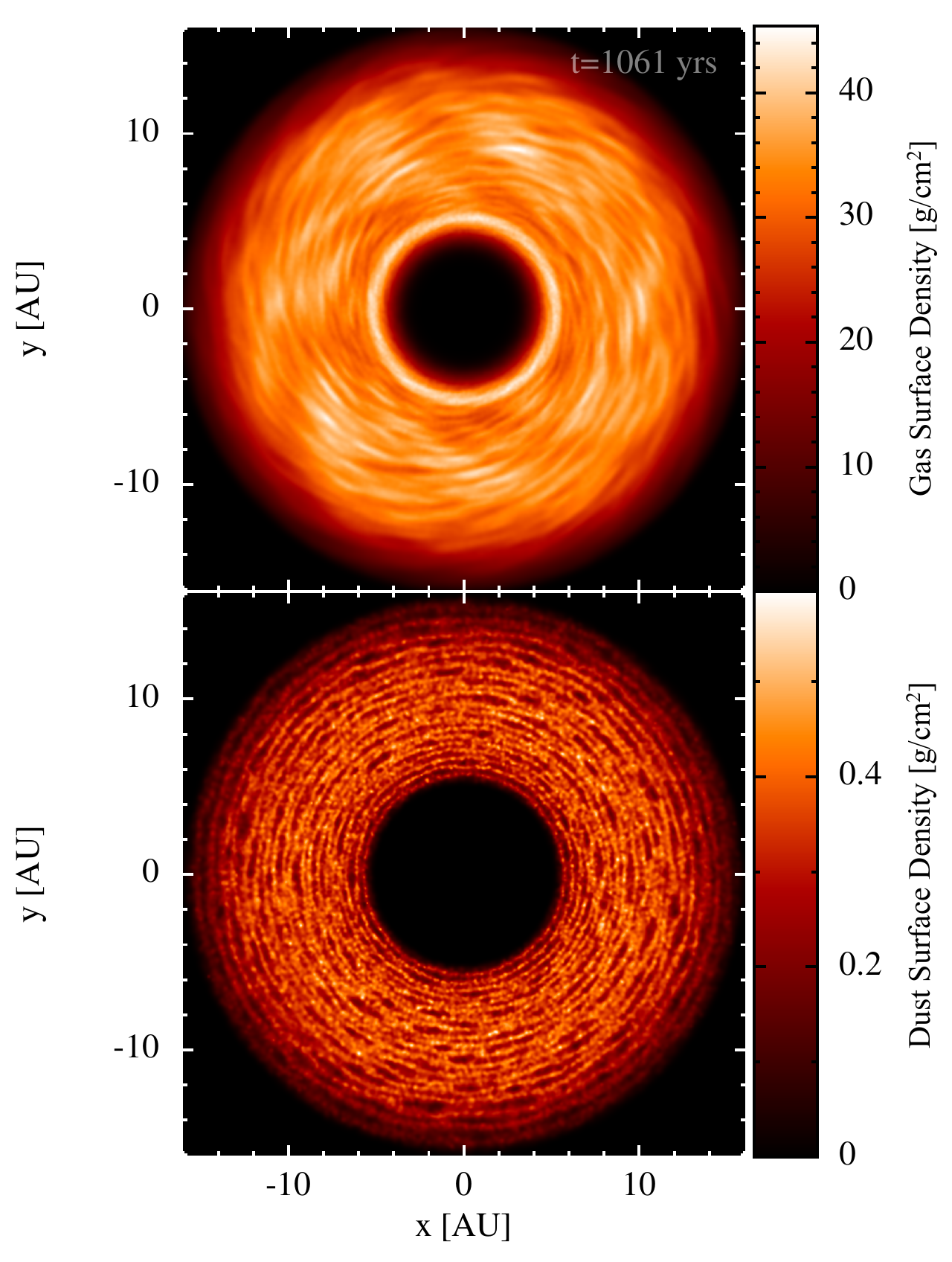}\vspace{-6pt}
\caption{Surface density renderings of the gas (top) and dust (bottom) after the instability has developed throughout the disc, for the same calculation as that depicted in Figs. 1 and 2. The same calculation as that shown in Figs. 1 and 2, but after the instability has developed throughout the disc.  The dust is conglomerated into concentric rings with peak densities located at the centres of the toroidal vortices within the dust layer. }
\label{fig:surface}
\end{figure}
\begin{figure} \centering\vspace{-12pt}
\includegraphics[width=80mm]{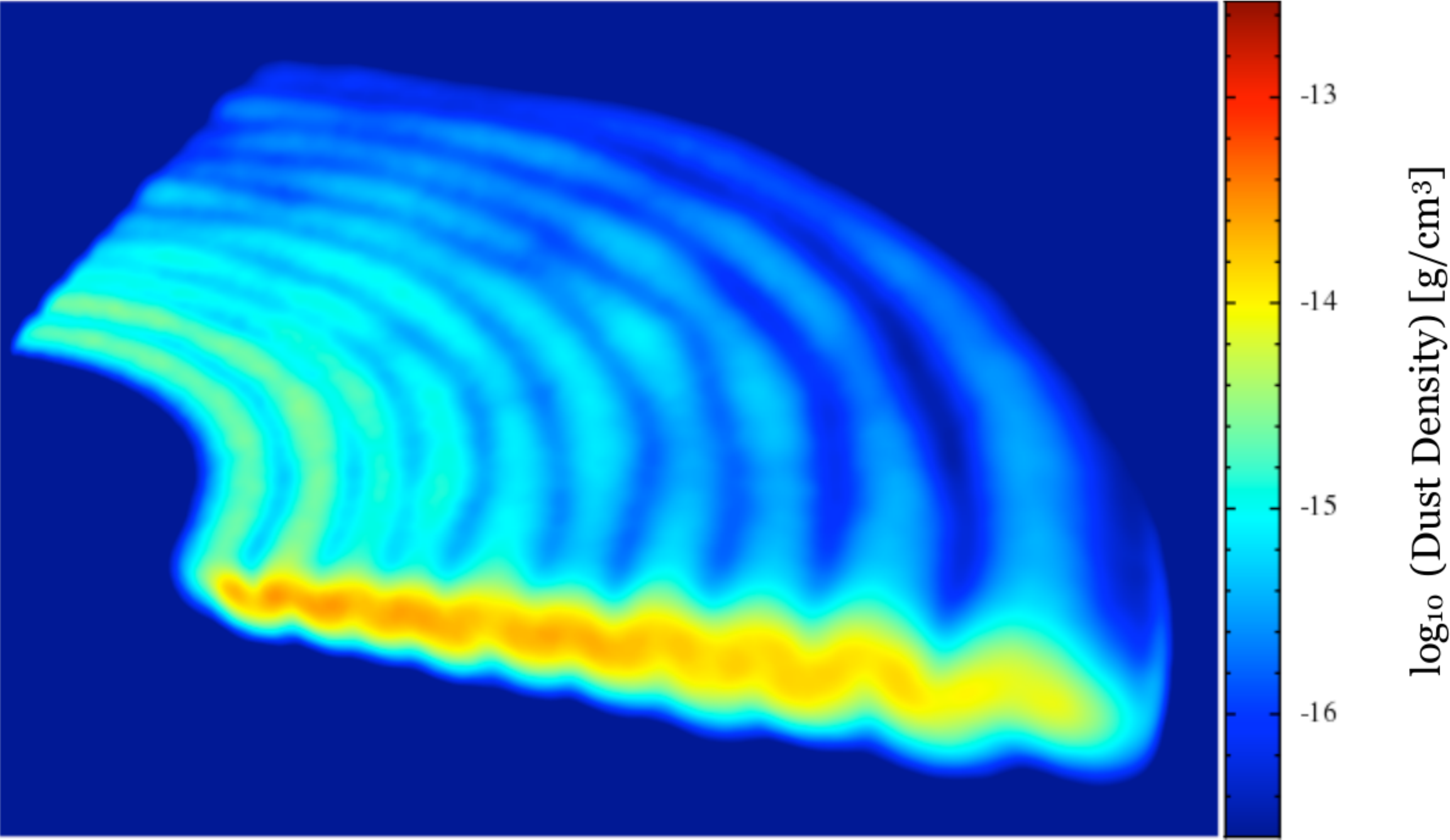}
\caption{A volume density rendering of the dust component whose surface density is shown in Fig. 3.}
\label{fig:volume}
\end{figure}

\vspace{-13pt}
\section{Numerical method}
\label{numerical}

To perform numerical simulations of dusty protoplanetary discs, we have used the three-dimensional SPH \citep{Lucy1977, GM77} code, {\tt sphNG} \citep{Aea11}, which has been used to study a variety of astrophysical fluid dynamical systems. For the simulations performed for this letter, the code uses two populations of SPH particle: gas and dust.  The latter only evolves under the action of gravity and a drag force.  In order to perform the time-integration of the dust-gas drag force, use the semi-implicit integration method of \cite{LB14}. 

The simulations performed followed the time evolution of locally-isothermal ($T \sim r^{-1}$ and independent of time) circumstellar gas/dust discs. The central star was modelled as a gravitational point mass, $M_{*} = 1~{\rm M}_{\odot}$, the mass of the gas component between the inner and outer radii was set to be $\approx 0.003 ~{\rm M}_\odot$ (similar to the minimum-mass Solar nebula), and the mass of the dust component was determined by the initial dust-to-gas ratio. The gaseous component of the disc had an outer radius of $r = 16.1$~AU, with an inner boundary at $r = 2.6$~AU, chosen to avoid having particles with excessively short integration time-steps. We modelled discs with vertical scale heights of $H/r=0.025$, 0.05, or 0.10 and surface density profiles $\Sigma \propto r^{-1/2}$.  Self-gravity was neglected.   We performed calculations with dust-to-gas ratios of $\epsilon = 0.001$, 0.01, or 0.1.  Calculations were typically performed with resolutions of $N_{\rm p} = 6\times 10^5$ SPH particles, but some cases were performed with $N_{\rm p} = 3\times 10^6$.  In each case,  approximately 1/3 of the particles were dust particles.  The initial dust-to-gas ratios were small enough to avoid significant perturbations of the gas hydrostatic equilibrium. After gas component relaxation, dust grains were introduced into the disc, following the gas density distribution, but with minimum and maximum orbital radii of $r = 5.2$ AU and $r = 15.6$ AU, respectively.  We performed simulations assuming spherical dust grains with radii of 1 $\mu$m, 0.1 mm, 1 mm, 1 cm, 1 m and 10 m, each with intrinsic dust grain densities of 3.0 g~cm$^{-3}$. 

Calculations performed at different numerical resolutions produced qualitatively similar results in terms of whether the instability developed or not, though the radial spacing of the vortices tended to decrease somewhat as the resolution was increased. We performed calculations using different SPH kernels: the M4 cubic, the M6 quintic \citep{MonLat1985}, and the C4 Wendland kernel \citep{Wendland1995}.  Very similar results were obtained with different kernels.  The figures in this letter were all produced from a calculation using $N_{\rm p} = 3\times 10^6$ and the M4 kernel.

\vspace{-12pt}
\section{Results and discussion}
\label{results}

To study the instability, we performed a series of three-dimensional smoothed particle hydrodynamic (SPH) simulations of dust settling in gaseous locally-isothermal Keplerian discs.   As a typical case, consider a gas disc with $H/r=0.05$ with a grain size of 1~mm and a dust-to-gas ratio of 0.01.  For radii of 5-10 AU, the dust settling times are of order a hundred years for such partially coupled grains.  The dust particles reach a stationary solution where their vertical velocity becomes negligible \citep*{NSH86}, producing a dust layer centred on the mid-plane with a width inversely proportional to their stopping time. As can be seen in Fig. \ref{fig:evolution}, soon after reaching the quasi-stationary phase, the disc starts to develop the baroclinic instability. Toroidal vortices develop within the dust layer (Fig. \ref{fig:vortices}), creating a sustained convective motion  (see the video at http://www.astro.ex.ac.uk/people/mbate/Animations/dust1.html for an illustration of the instability development). Dust particles collect at the centres of the vortices, so that looking down on the disc the dust conglomerates into rings (lower panel of Fig. \ref{fig:surface}).  

We find that the radial spacing of the vortices (and therefore the dust rings) is approximately equal to the thickness of the dust layer when the instability begins.  Therefore, the radial spacing is expected to increase for thicker and/or more massive gaseous discs, or smaller or less dense dust grains.  If the radial spacing between rings is large enough, they may be observable at sub-millimetre and millimetre wavelengths.  Furthermore, in Figs. \ref{fig:vortices} and \ref{fig:volume} it can be seen that the dust layer is corrugated, with the vertical peaks being out of phase with the density maxima of the dust rings.  Full radiative transfer modelling is beyond the scope of this letter, but we note that because of the vertical corrugation of the dust layer the rings may also be visible in scattered light at optical wavelengths or due to thermal shadowing at near-infrared wavelengths.  Indeed, it is tempting to speculate that the rings seen in the scattered light images of the TW Hya disc \citep{Debes2013} or the recent Atacama Large Millimetre Array (ALMA) observations of the HL Tau disc which displays several concentric rings \citep{Bea15} may be manifestations of this instability. Since self-gravity is neglected, the calculations are scale-free as long as dimensionless quantities are kept constant. For example, scaling the simulations to cover radii eight times larger (outer radius 128 AU -- the approximate radius of the HL Tau disc), to maintain the same dimensionless stopping time, either the gas mass of the disc must be increased by a factor of 64, or the size of the dust particles must be reduced the same factor (or a combination of the two).  Other mechanisms that may produce dust rings in discs include embedded protoplanets \citep[e.g.][]{PaaMel2004}, or the photoelectric instability proposed by \cite{LK13} in low density discs.

From our other calculations with different parameters, we find that the higher the dust-to-gas ratio, the more easily and quickly the instability develops.  However, we emphasise that the instability does not require large dust-to-gas ratios.  The exact behaviour depends on the stopping time, which in turn depends on the dust grain size and density, and gas density.  We find that grains with sizes smaller than 0.1~mm are tightly coupled to the gas, and the disc does not achieve a sufficient compositional gradient to drive the instability, at least over the timescales we have modelled.  Conversely, grains with sizes of $\gtrsim 10$~cm or more are too weakly coupled to the gas for it to act as a single fluid and drive the instability.   They rapidly settle into very thin layers near the disc mid-plane without any evidence of the instability. The exact cut off depends on the dust-to-gas ratio and the gas density. For our low-density discs with gas scale-heights of $H/r=0.10$, grains with sizes of 1~cm are too weakly coupled to the gas to drive the instability even with an initial dust-to-gas ratio of 0.1. With $H/r=0.05$, the instability develops with 1~cm grains and a dust-to-gas ratio of 0.1, but not 0.01.  With $H/r=0.025$, 1~cm grains drive instability for dust-to-gas ratios of 0.01 and higher.

These numerical results are in good agreement with the analytical model described above, since the closer to the mid-plane the dust is (i.e. the larger the grains, or lower the gas density), the harder it should be for the instability to develop. In equation \ref{eq:BV}, one can see that the contribution of the vertical gradient of the dust-to-gas ratio to the Brunt-V\"ais\"al\"a frequency is proportional to the product of the entropy gradient induced by it, and $g_{\rm z}$. Hence, for $z$ values close to the mid-plane, a very strong gradient of the dust-to-gas ratio will be necessary to compensate for the weakness of the gravitational field. Close to the mid-plane, the instability will not develop unless $\left| \partial \epsilon / \partial z \right| \gtrsim 1/z$ (equation \ref{eq:instability}).

In past studies, baroclinic instabilities in discs have been proposed to arise due to radial thermal entropy gradients \citep{KB03, Klahr04, PJS07}.  Not only is the driving mechanism completely different from the one we discuss in this letter, but radial entropy gradients produce vortices that are coplanar with the disc rather than toroidal.  Some recent local \citep{BS10} and global \citep{Kea13} simulations of dusty discs have investigated the streaming instability \citep{YG05, JY07a, JY07b} and found ring-like features. However, \cite{Kea13} exclude the vertical gravitational acceleration so there is no dust settling, they do not obtain complete rings, and they find that the perturbation wavelength decreases with increasing dust-to-gas ratio whereas we find no such dependence. \cite{BS10} included vertical gravity and studied the stability of very thin dust layers to Kelvin-Helmholtz and streaming instabilities.  Although they performed local simulations with much thinner dust layers than ours (of order 1/100 the gas scale height), in some of their calculations they find radial oscillations of the dust density similar to what we find here.  They ruled out Kelvin-Helmholtz instability as the driving mechanism due to the absence of structure in the azimuthal direction (as we do), and attributed the radial structures as `most likely due to' the streaming instability.  Instead, we propose that compositional entropy gradients due to the varying dust-to-gas ratio are crucial for the formation of these axisymmetric rings. The instability in Fig. \ref{fig:evolution} takes only $\approx 12$ orbital periods to develop, whereas the growth time of the streaming instability is expected to be $\approx 60-200$ orbital periods \citep{YG05} for these parameters ($\rho_{\rm dust}/\rho_{\rm gas}=0.02$, dimensionless stopping times $\tau_{\rm S} \approx 0.02$ at the mid-plane and $\tau_{\rm S} \approx 0.05$ at the edge of the dust layer), which we note is comparable to the dust settling time. Further investigation is certainly warranted.

\vspace{-12pt}
\section{Conclusions}
\label{conclusions}

We have shown that dusty protoplanetary discs may be unstable to a baroclinic instability that arises due to vertical compositional entropy gradients that are naturally produced during dust settling.  The baroclinic instability produces toroidal gas vortices that gather the dust into rings.  These rings are potentially observable at (sub-)millimetre wavelengths using ALMA, and with high-resolution telescopes operating at optical or near-infrared wavelengths. The susceptibility of dusty protoplanetary discs to develop toroidal vortices has implications for a wide range of disc processes that warrant future investigation, but are beyond the scope of this letter. It is unclear how the compositional baroclinic instability may interact with other instabilities in protoplanetary discs, such as the dust streaming instability, the thermal baroclinic instability, the magneto-rotational instability \citep{BH91}, or gravitational instability.  In the calculations presented in this letter, due to the axisymmetric initial conditions, the baroclinic instability produces global toroidal vortices.  Although compositional entropy gradients will be expected to develop in more structured discs, the manifestation of the resulting baroclinic instabilites may be different, and they may also alter angular momentum transport within the disc. Grains larger than $\sim 10$~cm cannot drive the instability because they are insufficiently coupled to the gas, however, their radial migration may be significantly altered by the gas vortices produced due to small grains.  If the migration of metre-sized objects can be slowed or stopped this may aid planet formation.  Finally, the long-term evolution of the vortices and dust rings needs to be examined, and whether or not the conglomeration of dust within the toroidal vortices may promote planetesimal growth, either by collisional agglomeration or gravitational instability.\\

\vspace{-24pt}
\section*{Acknowledgments}
The figures were created using SPLASH \citep{Pri07}, a SPH visualization tool publicly available at http://users.monash.edu.au/$\sim$dprice/splash.  

This work was supported by the STFC consolidated grant ST/J001627/1, and by the European Research Council under the European Community's Seventh Framework Programme (FP7/2007-2013 grant agreement no. 339248). This work used the DiRAC Complexity system, operated by the University of Leicester IT Services, which forms part of the STFC DiRAC HPC Facility (www.dirac.ac.uk). This equipment is funded by BIS National E-Infrastructure capital grant ST/K000373/1 and  STFC DiRAC Operations grant ST/K0003259/1. DiRAC is part of the National E-Infrastructure. This work also used the University of Exeter Supercomputer, a DiRAC Facility jointly funded by STFC, the Large Facilities Capital Fund of BIS and the University of Exeter.

\label{lastpage}
\end{document}